\documentclass[conference]{IEEEtran}
\IEEEoverridecommandlockouts
\usepackage{cite}
\usepackage{amsmath,amssymb,amsfonts}
\usepackage{algorithmic}
\usepackage{graphicx}
\usepackage{textcomp}
\usepackage{xcolor}
\usepackage{booktabs}
\usepackage{tabularx}
\usepackage{url}
\usepackage{multirow}

\usepackage{fancyhdr}
\usepackage{sc26repro}

\usepackage[caption=false,font=footnotesize]{subfig}
\def\BibTeX{{\rm B\kern-.05em{\sc i\kern-.025em b}\kern-.08em
    T\kern-.1667em\lower.7ex\hbox{E}\kern-.125emX}}
\usepackage{subfig}

\usepackage{fvextra}

\DefineVerbatimEnvironment{artifactcommand}{Verbatim}{
    fontsize=\footnotesize,
    fontfamily=tt,
    xleftmargin=1.2em,
    breaklines=true,
    breakanywhere=true,
    breaksymbolleft={},
    breaksymbolright={},
    baselinestretch=1
}

\begin{document}

\title{T-CCL: Resource Efficient and Performant Collective Communication using Tensor Memory Accelerator}

\author{
\IEEEauthorblockN{\makebox[0.30\textwidth][c]{Keyvan Dadashzadeh}}
\IEEEauthorblockA{
\textit{Chalmers University of Technology} \\
\textit{and University of Gothenburg}\\
Gothenburg, Sweden \\
keyvand@chalmers.se}
\and
\IEEEauthorblockN{\makebox[0.30\textwidth][c]{Yuehong Zhou}}
\IEEEauthorblockA{
\textit{Chalmers University of Technology} \\
\textit{and University of Gothenburg}\\
Gothenburg, Sweden \\
yuehong@student.chalmers.se}

\and[\hfill\mbox{}\par\mbox{}\hfill]

\IEEEauthorblockN{\makebox[0.30\textwidth][c]{Minyu Cui}}
\IEEEauthorblockA{
\textit{Linnaeus University}\\
Växjö, Sweden \\
minyu.cui@lnu.se}
\and
\IEEEauthorblockN{\makebox[0.30\textwidth][c]{Miquel Peric\`as}}
\IEEEauthorblockA{
\textit{Chalmers University of Technology}\\
\textit{and University of Gothenburg}\\
Gothenburg, Sweden \\
miquelp@chalmers.se}
}

\maketitle

\thispagestyle{fancy}
\lhead{}
\rhead{}
\chead{}
\lfoot{\footnotesize{
SC26 Workshops, November 15-20, 2026, Chicago, Illinois, USA
\newline 979-8-3195-1221-5/26/\$31.00 \copyright 2026 IEEE}}
\rfoot{}
\cfoot{}
\renewcommand{\headrulewidth}{0pt}
\renewcommand{\footrulewidth}{0pt}

\begin{abstract}

Large transformer-based models increasingly depend on multi-GPU execution,
which requires frequent collective communication among GPUs. Existing
communication libraries often rely on many GPU threads to achieve high
bandwidth or low latency, resulting in a large streaming multiprocessor
(SM)-side resource footprint. This footprint can limit the resources
available to other GPU work, particularly when communication and computation
execute concurrently. Thus, efficient collective communication should not
only achieve high collective performance but also reduce its SM-side resource usage.


This paper presents T-CCL, a resource-efficient collective communication library based on the Tensor Memory Accelerator (TMA) for intra-node communication. 
T-CCL offloads both data movement and reduction operations to TMA and executes each collective as a pipelined series of asynchronous TMA operations, reducing the SM resources required for collective communication while maintaining high bandwidth. 
Evaluated across AllReduce, AllGather, and ReduceScatter collectives, T-CCL outperforms NCCL by up to 2.4$\times$ with unrestricted communication resources and up to 3.42$\times$ under restricted resource budgets, remains competitive with NCCL's recent symmetric-memory kernels, and occupies the same or fewer SMs in profiled cases.
In a GEMM--collective overlap case study, switching the communication backend from NCCL to T-CCL raises the average operator-level speedup over a sequential baseline from 1.12$\times$ to 1.25$\times$ on two GPUs and from 1.04$\times$ to 1.14$\times$ on four GPUs, as T-CCL uses fewer SMs
for communication, leaving more SMs available to the overlapped GEMM.
%
Integrated into vLLM as a communication backend, T-CCL improves end-to-end inference throughput over vLLM's automatic backend dispatch by up to 1.31$\times$, outperforming it at every evaluated batch size on both the conversation and decode-heavy workloads. 

\end{abstract}


\section{Introduction}
The scale of modern transformer-based models often makes single-GPU execution impractical. Their parameters, activations, and intermediate tensors may exceed the memory available on one device.
Consequently, training and inference increasingly rely on execution across multiple GPUs.
Several parallelization strategies distribute model computation across multiple GPUs. Tensor parallelism is widely used and partitions the weights and computation of individual layers among the participating devices. Each GPU computes a partial result, which must then be combined or
redistributed through collective operations such as AllReduce,
ReduceScatter, and AllGather~\cite{megatron_bridge_parallelism,nccl}.
Because these collectives are invoked repeatedly within transformer blocks, they often lie on the critical execution path. Performance therefore depends on both computation efficiency and collective communication  efficiency.
The performance of collective operations depends on several system characteristics, including the number of participating GPUs, the interconnect topology, and the communication paths involved. These paths may remain within one node or may traverse links between nodes.
In this work, we focus on systems with multiple GPUs in one node, where each participating GPU has peer-to-peer access to one another's memory.

Starting with Hopper-generation GPUs, NVIDIA introduced the Tensor Memory Accelerator, dedicated hardware designed for asynchronous bulk movement of multidimensional tensor data between global memory (HBM) 
and shared memory (SMEM)~\cite{cuda_guide}. 
Conventional transfers were
commonly performed through thread-driven instructions, with many threads participating in address generation and data movement. Consequently, the achievable transfer throughput and resource cost depended on the number of threads assigned to the operation. 
In contrast, TMA allows a bulk transfer to be initiated by a single elected thread within a thread block (TB), after
which the transfer proceeds asynchronously through dedicated hardware. This
offloads address generation and data movement from GPU threads, allowing the
thread block to continue other computation while the transfer is in
flight~\cite{cuda_guide}.
Due to its asynchronous nature, TMA is attractive for collective
communication, where partial results must be transferred at high bandwidth
while keeping the SM resource footprint small.
In intra-node systems where GPUs have peer-to-peer access to one another's
memory, TMA can be used alongside thread-driven load/store instructions and
copy engines. Thread-driven communication is flexible and can directly
perform reductions, but it consumes SM resources, whereas copy engines move
data asynchronously but do not directly support element-wise reduction. TMA
supports both asynchronous bulk data movement and reduction operations. Based
on these capabilities, we make two observations that motivate its use as
a mechanism for intra-node collective communication~\cite{cuda_guide}.

\textbf{First,} TMA provides a resource-efficient mechanism for moving
data between global memory and shared memory, including
transfers to or from peer GPU memory. Conventional thread-driven collective
kernels require many threads to perform such data movement, whereas a TMA
transfer can be initiated by a single thread and proceeds asynchronously.
Therefore, TMA-based communication has the potential to sustain high bandwidth with fewer threads and CTAs, thereby reducing its SM resource footprint.
%

\textbf{Second,} beyond data movement, TMA can offload a substantial part
of the reduction work required by collective communication. 
Data movement is only one component of a collective; operations such as AllReduce and ReduceScatter must also combine values contributed by multiple GPUs through an element-wise reduction, such as summation~\cite{nccl}.
In conventional GPU kernels, GPU threads are responsible not only for moving data but also for performing these reductions, and contributions from multiple GPUs must be coordinated to avoid conflicting updates. Therefore, reductions are a major source of communication-side resource usage.  
%
TMA also supports asynchronous bulk reduction operations, including addition, minimum, and maximum. 
PTX defines these operations as asynchronous element-wise atomic
reductions, which can simplify the coordination of concurrent updates and
reduce the amount of thread-level work required for reduction~\cite{ptx_isa}.

Based on these observations, we present \textbf{T-CCL}, a TMA-based collective
communication library for intra-node communication. 
The goal of T-CCL is to deliver a better tradeoff between collective bandwidth and SM usage than NCCL~\cite{nccl}, particularly for resource-constrained execution in which communication shares the GPU resources with computation.
The contributions of this work are as follows:
\begin{itemize}
\item We design and implement T-CCL, an intra-node collective communication library that offloads both inter-GPU data movement and reduction to TMA. 
T-CCL supports AllReduce, ReduceScatter, and AllGather collectives.

\item We evaluate T-CCL against NCCL and NCCL's
symmetric-memory kernels, and show that T-CCL delivers comparable or
better performance across AllReduce, ReduceScatter, and AllGather, with
only a few small-message regressions, while using the same or fewer SM
resources.

\item In a first case study on compute--communication overlap, a widely adopted latency-hiding technique in which communication and computation compete for SM resources, 
we show that T-CCL's lower SM usage can translate into higher operator-level overlap speedup than NCCL-based overlap.

\item In a second case study, we integrate T-CCL into vLLM as a
communication backend and evaluate it under tensor-parallel inference,
showing that its communication-level gains translate into higher
end-to-end throughput, with the largest improvements on a decode-heavy
workload where collectives are invoked more frequently.
\end{itemize}


\section{Related work}
\label{sec:related}
Collective communication libraries (CCLs) provide the communication
backbone for distributed training and inference. Among well-known CCLs,
NCCL has become the de facto standard high-performance collective backend
for NVIDIA GPU systems. NCCL abstracts topology discovery, communication
setup, and the selection of algorithms, protocols, channels, and transport
paths~\cite{nccl,demystifying_nccl}. This design hides much of the
complexity of collective communication from users and allows NCCL to
operate efficiently across a wide range of systems. Following this design philosophy, 
T-CCL similarly hides low-level scheduling and communication setup from users.

Starting with version 2.27, NCCL introduced symmetric-memory kernels to
reduce latency and improve bandwidth. Recent NCCL releases also integrate
TMA into the load/store path of selected symmetric-memory kernels, making
NCCL one of the first widely used CCLs to adopt TMA for collective
communication. This TMA path is enabled by default only on Blackwell and
newer architectures. Unlike T-CCL, which accepts user-provided buffers, symmetric
NCCL requires communication buffers to be allocated using CUDA Virtual
Memory Management (VMM)-based allocators, complicating direct use with
framework-managed buffers such as PyTorch tensors.

While NCCL and T-CCL hide much of the communication process from users,
other systems expose more of it to provide greater programmability. As a
case in point, MSCCL allows users to define custom collective schedules
through a domain-specific language, while MSCCL++ further extends this
approach with more programmable abstractions and access to lower-level
communication primitives~\cite{msccl,mscclpp}. NVSHMEM similarly exposes
low-level communication primitives through a partitioned global address
space and GPU-initiated memory operations, enabling fine-grained
communication from device code~\cite{nvshmem}.

Beyond programmability and ease of use, some systems, more closely aligned
with the principles of T-CCL, focus on reducing the resource usage of
collective communication. To start with, ResCCL observes that many CCLs use
static CTA allocations and inefficient internal scheduling
mechanisms~\cite{res_ccl}. To address this, ResCCL schedules communication
primitives according to their data dependencies and supports flexible CTA
allocation to reduce pipeline bubbles, improve bandwidth utilization, and
lower SM usage. VCCL~\cite{vccl} takes a different approach to resource
efficiency by offloading point-to-point data movement that does not involve
reduction to CPU threads and GPU copy engines.

Finally, another category of work explores reducing communication overhead
through new hardware designs or co-design with emerging hardware
capabilities. For example, T3 proposes hardware support for tracking computation progress,
triggering communication, and performing data movement through DMA without
requiring additional GPU compute units~\cite{T3}. ParallelKittens, which builds on
ThunderKittens~\cite{thunderkittens}, demonstrates efficient data movement
and reduction using NVSwitch-supported communication
mechanisms~\cite{parallelkittens}. In contrast, T-CCL uses the TMA unit
already available within each GPU as its communication substrate, without
requiring additional interconnect-level or custom hardware support.

\section{System Design}
In this section, we introduce the main components of T-CCL and describe how it executes a collective operation, from the high-level interface down to the underlying data movement. T-CCL is organized as a set of high-level and low-level abstractions that enable collective operations to be executed efficiently. We first describe the programming model exposed by T-CCL and how users invoke collective operations through this interface. We then explain how each collective is translated into a set of logical transfer and reduction operations, and how these logical operations are lowered into physical transfer tasks specific to each GPU. Finally, we present the TMA engine, which serves as the main execution component
and efficiently carries out these physical tasks by issuing TMA instructions
in a pipelined manner.

\begin{figure*}[t]
    \centering
    \subfloat[Shard-in phase. The owner of a shard receives the corresponding
        contributions from the other GPUs and combines them to produce the final
        value of that shard locally. In this example, GPU~0 owns shard~0, receives
        the shard~0 contributions from the other ranks, and reduces them locally to
        produce the final value.]{
        \includegraphics[width=0.48\textwidth]{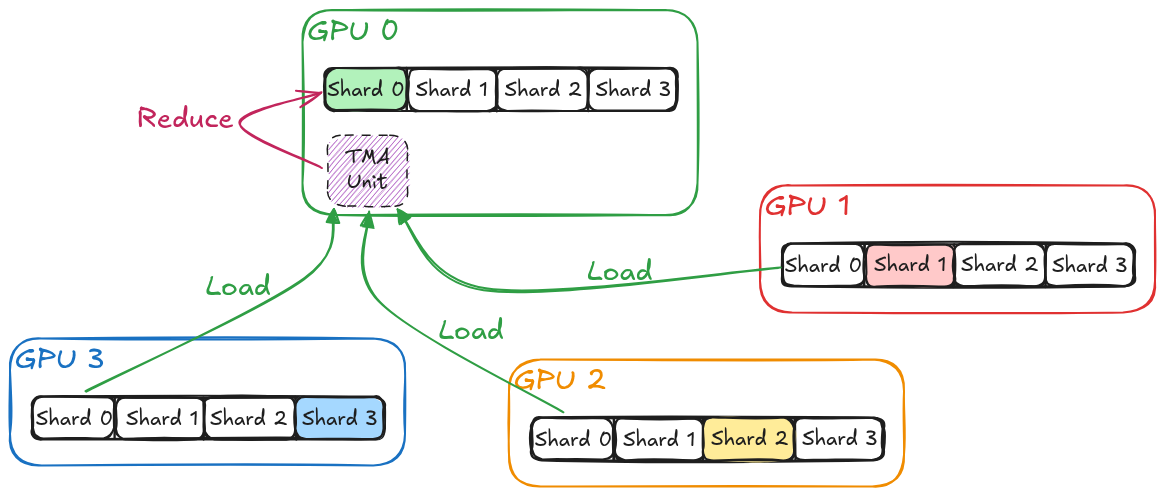}
        \label{fig:shard-in}
    }
    \hfill
    \subfloat[Shard-out phase. The owner of a shard is responsible for propagating that
        shard to the other GPUs that require it. In this example, GPU~0 owns
        shard~0 and propagates it to the other ranks.]{
        \includegraphics[width=0.48\textwidth]{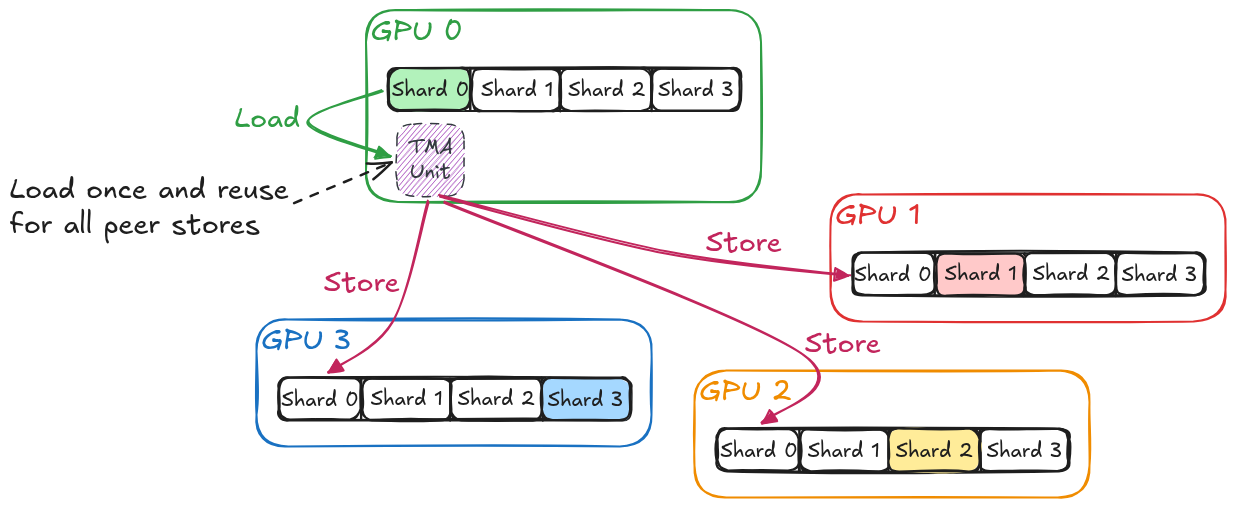}
        \label{fig:shard-out}
    }
    \caption[Shard ownership and lifecycle in T-CCL for a four-GPU group.]{
    Shard ownership and lifecycle in T-CCL for a four-GPU group. Shard colors
    indicate ownership, with each shard colored according to its owning GPU.
    Figure~\protect\subref{fig:shard-in} shows GPU~0 receiving and reducing
    the shard~0 contributions from the other ranks.
    Figure~\protect\subref{fig:shard-out} shows GPU~0 propagating shard~0
    to the other ranks.}
    \label{fig:shard-ownership}
\end{figure*}

\subsection{Programming Model}

T-CCL organizes the participating GPUs into a communication
\textit{Group}, where each GPU is assigned a unique rank. T-CCL supports
AllReduce, ReduceScatter, and AllGather in both single-process and
multiprocess execution modes. The group-creation interface differs between 
the two modes. In single-process mode, one process creates a group containing
all participating GPUs. In multiprocess mode, each process creates a local
view of the communication group. To synchronize the processes and exchange
the information required to access peer buffers, T-CCL adopts the
interprocess communication utilities from
ThunderKittens~\cite{thunderkittens}.
After group creation, both execution modes use the same collective
interfaces. These interfaces follow the collective semantics of
NCCL~\cite{nccl}, which simplifies the integration of T-CCL into
applications built around NCCL. Each rank passes only its local buffer,
while T-CCL internally handles access to the corresponding buffers on the
other participating ranks.

T-CCL supports both in-place collectives, which reuse the input buffer for
the output, and out-of-place collectives, which use separate input and
output buffers. T-CCL favors in-place execution because it can reduce the
number of transfer tasks and, in multiprocess mode, avoids exchanging
information for separate output buffers.

\subsection{Task-Based Collective Planning}

\subsubsection{Shard Ownership}
For a collective with $P$ participating ranks, T-CCL partitions the
logical tensor into $P$ shards and assigns one shard to each rank. The
owner of a shard is responsible for producing its final value and, when
required, propagating that value to the other ranks.
Fig.~\ref{fig:shard-ownership} illustrates this design for shard~0 in a
four-GPU group. In the \textit{shard-in} phase shown in
Fig.~\ref{fig:shard-ownership}\subref{fig:shard-in}, the contributions corresponding to shard~0 are
transferred from the other GPUs to GPU~0 and reduced there. In the
\textit{shard-out} phase shown in Fig.~\ref{fig:shard-ownership}\subref{fig:shard-out}, GPU~0
propagates the final value of shard~0 to the other GPUs. In both phases,
GPU~0 owns shard~0 and drives the associated operations because
peer-to-peer memory access allows it to access the required data on the
other GPUs without their active participation.

T-CCL prioritizes reduction into the owner's local shard rather than
directly into peer memory. Cross-device reduction is more expensive than
loading data from peer memory and reducing it locally. It also requires
stricter atomic guarantees because multiple GPUs may
concurrently target the same memory region. By performing reduction in the
\textit{shard-in} phase and propagation in the \textit{shard-out} phase,
T-CCL reduces the need for fine-grained cross-device coordination.
%
As a result, synchronization is mainly required at the boundaries of the
collective: at the beginning, to ensure that the input buffers are ready,
and at the end, to ensure that all assigned work has completed. Because 
the number of these synchronization points does not grow with the
message size, the synchronization overhead does not scale with the message size.

\subsubsection{Logical Transfer Tasks}

T-CCL uses a \textit{task} as the basic unit of communication and execution
between GPUs. The component responsible for generating these tasks is the
\textit{planner}. The planner tracks the location and ownership of each shard
and generates the logical copy, reduction, and synchronization tasks required
to produce the final value of every shard. 
Each logical task specifies the operation, the executing rank, the
source and destination regions, and its dependencies
on other tasks. The planner runs on only one rank, typically the rank with
the lowest identifier in the group. Once the logical plan is ready, it is
shared with the other ranks through pre-allocated buffers.

Each rank independently lowers its assigned logical tasks into physical
transfer tasks, resolving details such as memory addresses and the transfer
mechanism used to execute each task. This separation is
important because the planner requires a global view of the topology to
construct an efficient collective execution plan, whereas physical-task
generation deals with the constraints and efficient execution of that plan
on an individual GPU. The design therefore enables optimization at both the
topology level and the device-execution level. In this work, all evaluations use symmetric GPU topologies and the same
communication resource budget for every rank. We therefore use a simple
planner that generates the same set of tasks for all ranks.

\subsection{TMA Transfer Engine}

\begin{figure*}[t]
    \centering
    \includegraphics[width=0.82\textwidth]{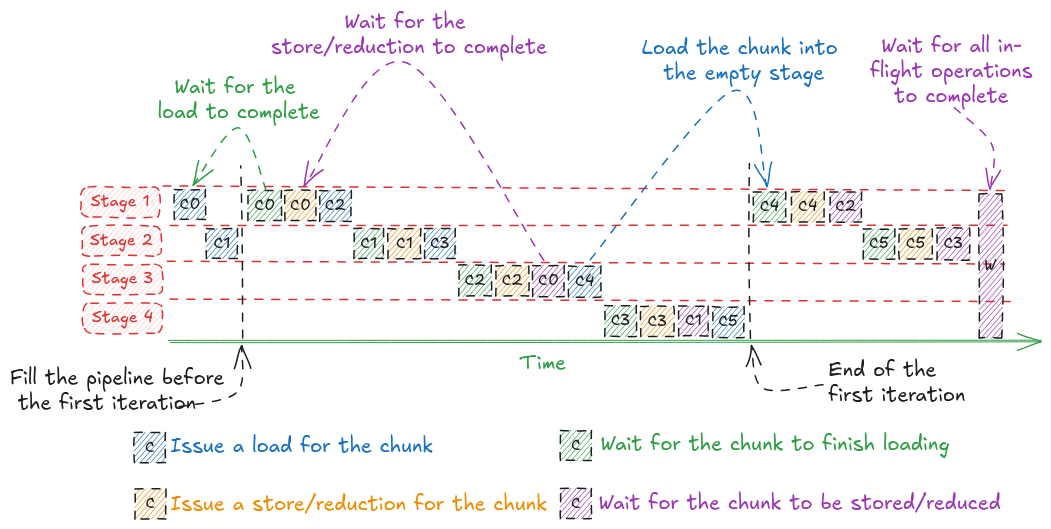}
    \caption{Pipeline execution of T-CCL's TMA engine for six chunks using four stages
allocated in shared memory.
    }
    \label{fig:tma-pipeline-example}
\end{figure*}

\subsubsection{Chunk and Window Decomposition}
Before execution, each logical task is translated into one or more physical
tasks. A physical task consists of a series of \textit{windows} that must be
processed, and each window is further divided into fixed-size chunks. A
\textit{chunk} is the basic unit processed by the TMA engine at each step, whereas a
window groups a series of chunks and is used to partition larger transfers. 
A window also provides a natural granularity for tracking the progress of
a group of chunks. For future work, we plan to support window-level
completion signaling to pipeline dependent reduction and copy operations
more closely.

Unlike logical-task generation, where the logical plan is generated once and
shared among the ranks, physical tasks are generated and optimized
independently by each rank according to its local constraints and
capabilities. T-CCL applies several optimizations to the resulting physical
tasks. We describe three of these prominent optimizations.

First, T-CCL applies \textit{fan-out}, which combines propagation tasks that read the same source region but write to different peer GPUs.
As illustrated in Fig.~\ref{fig:shard-ownership}\subref{fig:shard-out}, GPU~0 loads a chunk of its
local shard~0 into shared memory once and reuses it for three stores, one to each of the
other GPUs. Without fan-out, each destination would require a separate
load. Thus, in this example, fan-out reduces the number of loads from three to one
without changing the number of stores.

Second, T-CCL assigns loads from different peer GPUs to separate
CTAs. As described in the \textit{shard-in} phase, contributions from the
other GPUs are loaded and reduced into the owner's local shard. To prevent
contention on one peer link from delaying transfers from other peers and
leaving the TMA hardware underutilized, T-CCL schedules the load-and-reduction
work associated with different peers on separate CTAs. For example, in
Fig.~\ref{fig:shard-ownership}\subref{fig:shard-in}, loads from different peer GPUs are executed by
different CTAs.

Third, because TMA is primarily designed for bulk data movement,
additional care is required to preserve its performance benefits for small
transfers. In particular, when the transfer size is small enough to fit
within the shared-memory capacity available to a single CTA, the TMA engine assigns the
entire transfer to one CTA. This avoids distributing a small amount
of work across multiple CTAs and reduces the associated scheduling and
coordination overhead.

\subsubsection{Pipelined Copy and Reduction}
T-CCL allocates a staging buffer in shared memory (SMEM) and divides it into
multiple stages. If the pipeline depth is $D$ and the chunk size is $C$, then
the total staging-buffer size is $D \times C$, which must fit within the
shared-memory capacity available to the executing CTA. For instance, H100 provides
228\,KB of shared memory per SM and allows up to 227\,KB to be addressed by
a single thread block~\cite{hopper_tuning_guide}.
Both the number of stages and the chunk size are
configurable. In our design, we logically use half of the stages,
$\frac{D}{2}$, for chunks whose loads are in flight and the other half for
chunks whose store or reduction operations are in flight. This division is
motivated by the different completion mechanisms used by TMA. Load completion is tracked through barriers, 
allowing the executor to wait
for a specific load. Store and reduction operations, in contrast, are
batched into asynchronous groups using a commit operation. These groups are
tracked in a FIFO-like order: waiting until a group completes also requires
all earlier committed groups to complete. Thus, the executor cannot wait
for an arbitrary store or reduction independently, as it can for a
load operation~\cite{ptx_isa}.

When the pipeline is initially empty, T-CCL first issues loads for
$\frac{D}{2}$ chunks to fill the pipeline. Execution then proceeds
iteratively over the stages. For each stage, T-CCL waits for the previously
issued load to complete and then issues the corresponding store or reduction.
Before issuing the next load into the stage that is $\frac{D}{2}$ positions
ahead, T-CCL waits for the previous store or reduction using that stage to
complete. This allows later chunks to be loaded while
earlier chunks are being stored or reduced. After all chunks have been issued,
T-CCL waits for the remaining in-flight operations to complete and drains the
pipeline. Fig.~\ref{fig:tma-pipeline-example} illustrates this procedure for
six chunks, including the pipeline-fill, steady-state, and drain phases.

\subsubsection{Launch Configuration Tuning}
Simply assigning more CTAs does not necessarily improve the performance of
the TMA engine. Efficient execution instead requires a balance between the
number of CTAs and the amount of transfer work assigned to each CTA, so that
the pipeline remains sufficiently utilized while asynchronous operations are
in flight. T-CCL therefore uses a simple offline tuning process. For each
collective, it evaluates different CTA configurations across various message
sizes, ranks the measured configurations by execution time, and stores the
ranked configurations in a lookup table indexed by collective operation and
message size. 
When a collective is invoked, each rank selects the entry for the closest
measured message size and uses the highest-ranked configuration that fits
the available CTA budget. This allows the TMA engine to select a
high-performing CTA configuration for a given collective and message size while remaining
within the available CTA budget.

\section{Evaluation}
\begin{figure*}[t]
    \centering
    \includegraphics[
        width=0.98\textwidth,
        keepaspectratio
    ]{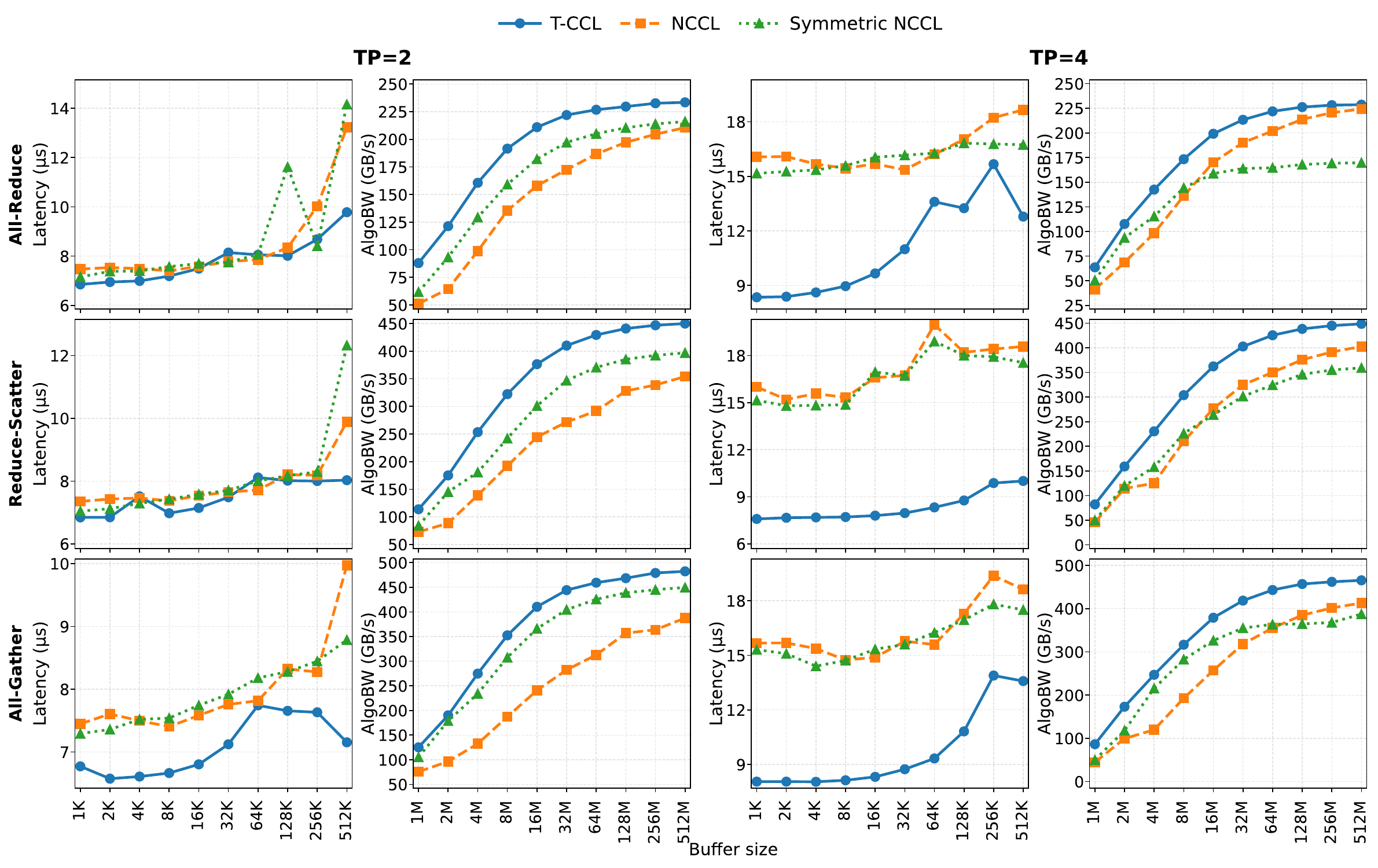}
    \caption{Collective-operation performance under unrestricted CTA budgets on
    the H100 and GH200 platforms. Rows correspond to AllReduce,
    ReduceScatter, and AllGather. The first two columns report latency and
    bandwidth for TP=2, while the last two columns report latency and
    bandwidth for TP=4.}
    \label{fig:collective-unrestricted}
\end{figure*}

\begin{table*}[t]
    \centering
    \caption{Average active SM usage and T-CCL speedup over NCCL configurations
    for AllGather and ReduceScatter on the GH200 platform. TMA NCCL denotes
    NCCL with TMA enabled.}
    \label{tab:sm-usage-speedup}
    \scriptsize
    \setlength{\tabcolsep}{2.5pt}
    \renewcommand{\arraystretch}{0.90}
    \begin{tabular}{llcccc|ccc}
        \toprule
        & & \multicolumn{4}{c|}{Average Active SMs}
        & \multicolumn{3}{c}{T-CCL Speedup ($\times$)} \\
        \cmidrule(lr){3-6}
        \cmidrule(lr){7-9}
        Collective & Size
        & T-CCL & NCCL & Sym.\ NCCL & TMA NCCL
        & NCCL & Sym.\ NCCL & TMA NCCL \\
        \midrule
        \multirow{6}{*}{AllGather}
        & 128\,KB & 1.32 & 3.30  & 1.98 & 1.65 & 1.53 & 1.59 & 1.98 \\
        & 512\,KB & 1.32 & 8.25  & 1.65 & 1.65 & 1.13 & 1.02 & 1.23 \\
        & 2\,MB   & 2.64 & 11.22 & 2.97 & 3.63 & 1.17 & 1.05 & 1.05 \\
        & 8\,MB   & 6.60 & 21.78 & 7.26 & 7.92 & 1.60 & 1.06 & 1.49 \\
        & 32\,MB  & 6.60 & 23.43 & 7.92 & 7.92 & 1.26 & 1.08 & 1.28 \\
        & 128\,MB & 6.60 & 23.76 & 9.24 & 9.24 & 1.15 & 1.13 & 1.20 \\
        \midrule
        \multirow{6}{*}{ReduceScatter}
        & 128\,KB & 1.32 & 3.30  & 1.98 & 1.65 & 1.21 & 1.25 & 1.55 \\
        & 512\,KB & 2.31 & 8.25  & 2.64 & 2.31 & 1.45 & 1.41 & 1.67 \\
        & 2\,MB   & 2.64 & 11.22 & 3.30 & 2.64 & 1.17 & 1.05 & 1.27 \\
        & 8\,MB   & 7.92 & 21.78 & 7.92 & 8.91 & 1.40 & 1.30 & 1.50 \\
        & 32\,MB  & 7.92 & 23.10 & 9.24 & 7.92 & 1.20 & 1.28 & 1.10 \\
        & 128\,MB & 9.24 & 23.76 & 9.24 & 9.24 & 1.11 & 1.20 & 1.02 \\
        \bottomrule
    \end{tabular}
\end{table*}

\subsection{Experimental Setup}\label{setup}
\subsubsection{Evaluation Settings}
We evaluate T-CCL in three settings. First, we evaluate T-CCL as a
standalone communication backend and compare its collective-operation
performance, with and without resource constraints, against NCCL and
symmetric NCCL. Second, we study how T-CCL improves
compute--communication overlap by maintaining high communication
bandwidth under limited resource budgets. Finally, we integrate T-CCL
with vLLM and evaluate its end-to-end inference performance under two
workloads.

\begin{table}[t]
\caption{Experimental environments.}
\label{tab:experimental-environments}
\centering
\scriptsize
\setlength{\tabcolsep}{3pt}
\renewcommand{\arraystretch}{1.1}
\resizebox{\columnwidth}{!}{
\begin{tabular}{lcc}
\hline
\textbf{Component} &
\textbf{H100 platform} &
\textbf{GH200 platform} \\
\hline
GPU &
$2 \times$ H100 NVL &
$4 \times$ GH200 120\,GB \\

CPU &
$2 \times$ AMD EPYC 9354 &
$4 \times$ ARM Neoverse-V2 \\

GPU interconnect &
NV12 &
NV6 
\\

CUDA toolkit &
13.3 &
13.0 \\
\hline
\end{tabular}
}
\end{table}

\subsubsection{Experimental Environment}
Unless otherwise stated, we conduct all two-GPU experiments on a system
with two H100 NVL GPUs (referred to as the \textit{H100} platform) and
all four-GPU experiments on a system with four GH200 GPUs (referred to as
the \textit{GH200} platform). The two GPUs in the H100 platform are
connected via NV12, whereas each pair of GPUs in the GH200 platform is
connected via NV6. Table~\ref{tab:experimental-environments} summarizes
the key system configurations of the evaluated platforms.

\subsubsection{Baselines} We compare T-CCL with standard NCCL and symmetric NCCL. Both are based
on NCCL version 2.28, which includes symmetric-memory kernels designed to
reduce synchronization overhead~\cite{nccl}.  Throughout this section,
\textit{symmetric NCCL} refers to the NCCL symmetric-memory kernel path. 
We include symmetric NCCL as a direct baseline only in the standalone collective evaluation, where the communication buffers can be allocated and registered specifically for this purpose. 
In the other evaluations, the buffers are provided by external components, and using symmetric NCCL directly would require additional data copies.  

For the LLM inference experiments, we use vLLM version 0.25.0.
Unless otherwise stated, we use the same T-CCL configuration across all
experiments. We set the chunk size to 8\,KB and the pipeline depth to 16,
resulting in a total shared-memory staging buffer of 128\,KB. Additionally,
because we do not use signaling between different communication phases, such
as reduction and copying, we set the window-to-chunk ratio to one, meaning
that each window consists of a single chunk. We enable T-CCL's offline
launch-configuration tuning in all experiments.

\begin{table}[t]
    \centering
    \caption{T-CCL speedup ranges over NCCL and symmetric NCCL across latency
    and bandwidth measurements under unrestricted CTA budgets.}
    \label{tab:collective-speedup-unrestricted}
    \scriptsize
    \setlength{\tabcolsep}{3pt}
    \renewcommand{\arraystretch}{1.05}

    \begin{tabular}{llcc}
        \toprule
        Platform & Collective & vs.\ NCCL & vs.\ Sym.\ NCCL \\
        \midrule
        \multirow{3}{*}{H100}
        & AllReduce     & 0.96--1.88$\times$ & 0.95--1.45$\times$ \\
        & ReduceScatter & 0.95--1.98$\times$ & 0.97--1.53$\times$ \\
        & AllGather     & 1.01--2.07$\times$ & 1.06--1.23$\times$ \\
        \midrule
        \multirow{3}{*}{GH200}
        & AllReduce     & 1.02--1.92$\times$ & 1.07--1.82$\times$ \\
        & ReduceScatter & 1.11--2.40$\times$ & 1.25--2.27$\times$ \\
        & AllGather     & 1.13--2.06$\times$ & 1.12--1.90$\times$ \\
        \bottomrule
    \end{tabular}
\end{table}

\subsection{Collective Operation Performance}

To evaluate the collective-operation performance of T-CCL, we organize the
results into two message-size ranges and report the metric most appropriate
for each. Small messages are especially relevant to latency-sensitive inference
phases such as decoding, whereas larger transfers arise in both training
and inference, including gradient exchange and prompt processing ~\cite{mscclpp}. 
The first range contains small messages
from \textit{1\,KB} to \textit{512\,KB}, for which we report collective
latency. The second range contains large messages from \textit{1\,MB} to
\textit{512\,MB}, for which we report the bandwidth achieved by the collective operations.
For each message size, we perform
50 warm-up iterations to reach
steady-state execution, followed by 250 timed iterations. We report the average execution time per iteration over these 250 iterations. 
Similar to the methodology used by NVIDIA's NCCL Tests~\cite{nccl-tests},
we vary the memory addresses used across consecutive iterations to
reduce cache-residency effects.
For bandwidth measurements, we use algorithmic bandwidth (denoted as AlgoBW), which is calculated by dividing the message size by the collective execution time.

We evaluate T-CCL, NCCL, and symmetric NCCL under two resource settings. In the first
setting, all three configurations are allowed to use enough CTAs to achieve their 
best communication performance. In the second setting, we impose
restricted CTA budgets, limiting communication to 9 CTAs on the GH200
platform and 8 CTAs on the H100 platform. We select these limits because, in our measurements, T-CCL already 
approaches bandwidth saturation
at 8 CTAs on H100 and 9 CTAs on GH200.


\paragraph{Unrestricted CTA Budgets}

Fig.~\ref{fig:collective-unrestricted} shows the performance of T-CCL,
NCCL, and symmetric NCCL under unrestricted CTA budgets on both the GH200
platform (TP=4) and the H100 platform (TP=2).
Table~\ref{tab:collective-speedup-unrestricted} summarizes the minimum and
maximum T-CCL speedups over NCCL and symmetric NCCL for each collective.
Overall, T-CCL outperforms both baselines across most evaluated message
sizes. The few regressions are small and occur primarily for small-message
cases on the H100 platform.

Symmetric NCCL, however, shows mixed results compared with standard NCCL.
We observe that symmetric NCCL uses a conservative CTA configuration for
AllReduce. In particular, it uses at most eight CTAs even as the message
size increases. As a result, it can perform worse than standard NCCL,
which scales its CTA usage more aggressively with message size and uses
more of the available communication resources. The same behavior is not
observed for ReduceScatter and AllGather, for which symmetric NCCL also
increases its CTA usage as the message size grows.

\paragraph{SM resource usage}
Although CTA count does not directly determine SM usage, it bounds the
number of SMs that a communication kernel can occupy because multiple CTAs
may be scheduled on the same SM. We therefore directly profile the average
SM usage of T-CCL, NCCL, symmetric NCCL, and NCCL with TMA enabled (denoted as \textit{TMA NCCL}) on the GH200 platform. While the standard and
symmetric NCCL baselines use NCCL~2.28, for \textit{TMA NCCL} we use
NCCL~2.30.7 and apply a small patch to enable its TMA path on the Hopper
architecture. We profile six message
sizes for both AllGather and ReduceScatter, including two small messages
and four large messages. Using Nsight Systems~\cite{nsight_systems}, we measure the average
number of active SMs over 1,000 iterations.
Table~\ref{tab:sm-usage-speedup} shows that T-CCL uses fewer than or the
same number of SMs as the other configurations while generally achieving
higher performance. 

Another notable observation is that \textit{TMA NCCL} performs worse than
T-CCL in all evaluated cases and is also slower than symmetric
NCCL in most experiments.
We attribute this result to three aspects of NCCL's current use of TMA in collective operations. 
\textbf{First,} NCCL primarily uses TMA for loading and storing data, while
reductions continue to be performed by a large number of GPU threads.
Because reduction is one of the most demanding components of collective
communication, using TMA only for data movement leaves much of its
potential benefit unexploited.
\textbf{Second,} NCCL waits for each TMA load or store to complete before
continuing. T-CCL instead pipelines multiple TMA operations to hide their
latency and improve throughput.
\textbf{Third,} NCCL processes loads from multiple peers within a single
CTA. Progress can therefore be limited by the slowest or most contended
peer link, leaving the TMA hardware underutilized. T-CCL assigns work
associated with different peers to separate CTAs, allowing transfers from
uncontended peers to proceed independently.

\begin{figure*}[t]
    \centering
    \includegraphics[
        width=0.98\textwidth,
        keepaspectratio
    ]{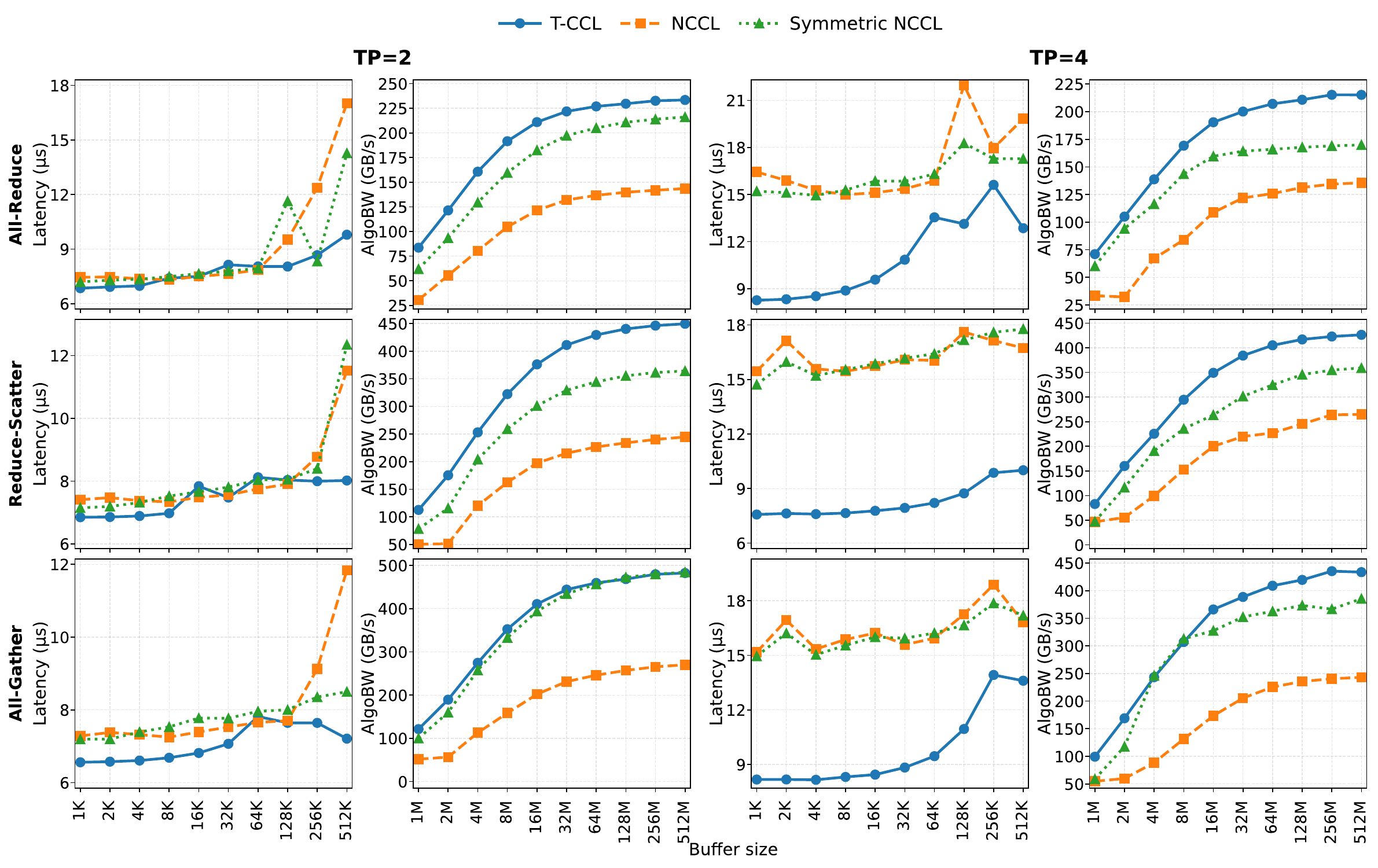}
    \caption{Collective-operation performance under restricted CTA budgets,
    with communication limited to 8 CTAs on the H100 platform and 9 CTAs on
    the GH200 platform. Rows correspond to AllReduce, ReduceScatter, and
    AllGather. The first two columns report latency and bandwidth for TP=2,
    while the last two columns report latency and bandwidth for TP=4.}
    \label{fig:collective-restricted}
\end{figure*}

\begin{table}[t]
    \centering
    \caption{T-CCL speedup ranges over NCCL and symmetric NCCL across latency
    and bandwidth measurements under restricted CTA budgets.}
    \label{tab:collective-speedup-restricted}
    \scriptsize
    \setlength{\tabcolsep}{3pt}
    \renewcommand{\arraystretch}{1.05}
    \begin{tabular}{llcc}
        \toprule
        Platform & Collective & vs.\ NCCL & vs.\ Sym.\ NCCL \\
        \midrule
        \multirow{3}{*}{H100}
        & AllReduce     & 0.94--2.74$\times$ & 0.96--1.46$\times$ \\
        & ReduceScatter & 0.95--3.42$\times$ & 0.98--1.54$\times$ \\
        & AllGather     & 0.98--3.35$\times$ & 0.99--1.21$\times$ \\
        \midrule
        \multirow{3}{*}{GH200}
        & AllReduce     & 1.15--3.26$\times$ & 1.11--1.84$\times$ \\
        & ReduceScatter & 1.60--2.88$\times$ & 1.18--2.09$\times$ \\
        & AllGather     & 1.24--2.83$\times$ & 0.98--1.98$\times$ \\
        \bottomrule
    \end{tabular}
\end{table}

\paragraph{Restricted CTA Budgets}
Fig.~\ref{fig:collective-restricted} shows the performance of T-CCL,
NCCL, and symmetric NCCL under restricted CTA budgets of 8 CTAs on the
H100 platform and 9 CTAs on the GH200 platform. Under these constraints,
T-CCL maintains higher bandwidth than both NCCL configurations across the
evaluated large-message sizes. Table~\ref{tab:collective-speedup-restricted}
summarizes the minimum and maximum T-CCL speedups over NCCL and symmetric
NCCL for each collective. Overall, T-CCL outperforms both baselines across
most evaluated message sizes, with only a few small regressions for
small-message cases on the H100 platform.

Symmetric NCCL also retains its performance better than standard NCCL under
restricted CTA budgets, consistent with the conservative CTA usage we
observe in the unrestricted setting. Consequently, restricting the CTA
budget affects standard NCCL more severely.

\begin{figure}[t]
    \centering
    \includegraphics[
        width=0.65\columnwidth,
        keepaspectratio
    ]
    {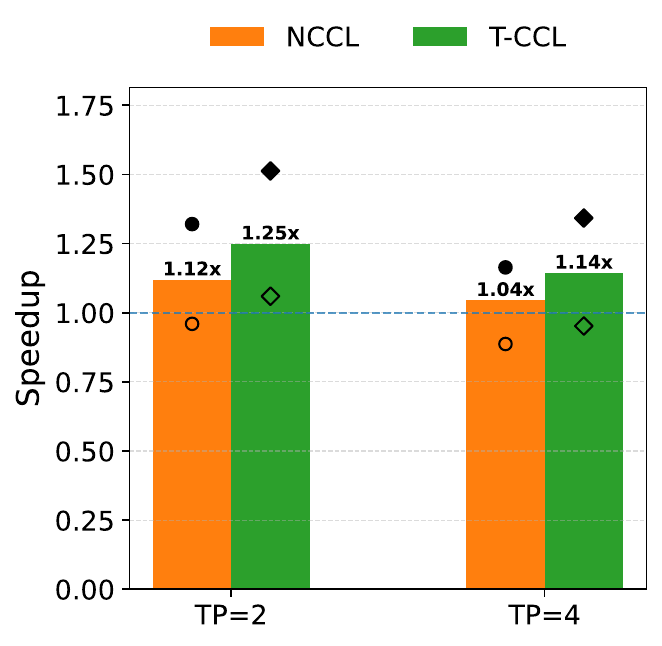}
    \caption{Average speedup of FlashOverlap using NCCL and T-CCL for
    TP=2 and TP=4. For each TP configuration, speedup is measured relative
    to its corresponding sequential cuBLAS--NCCL baseline. Bars show the
    average across 24 GEMM shapes, while the markers indicate the minimum
    and maximum speedups.}
    \label{fig:operator-overlap-speedup}
\end{figure}

\begin{figure*}[t]
    \centering
    \subfloat[Conversation workload: 1024 input tokens and 128 output
    tokens.]{
        \includegraphics[width=0.48\textwidth]
        {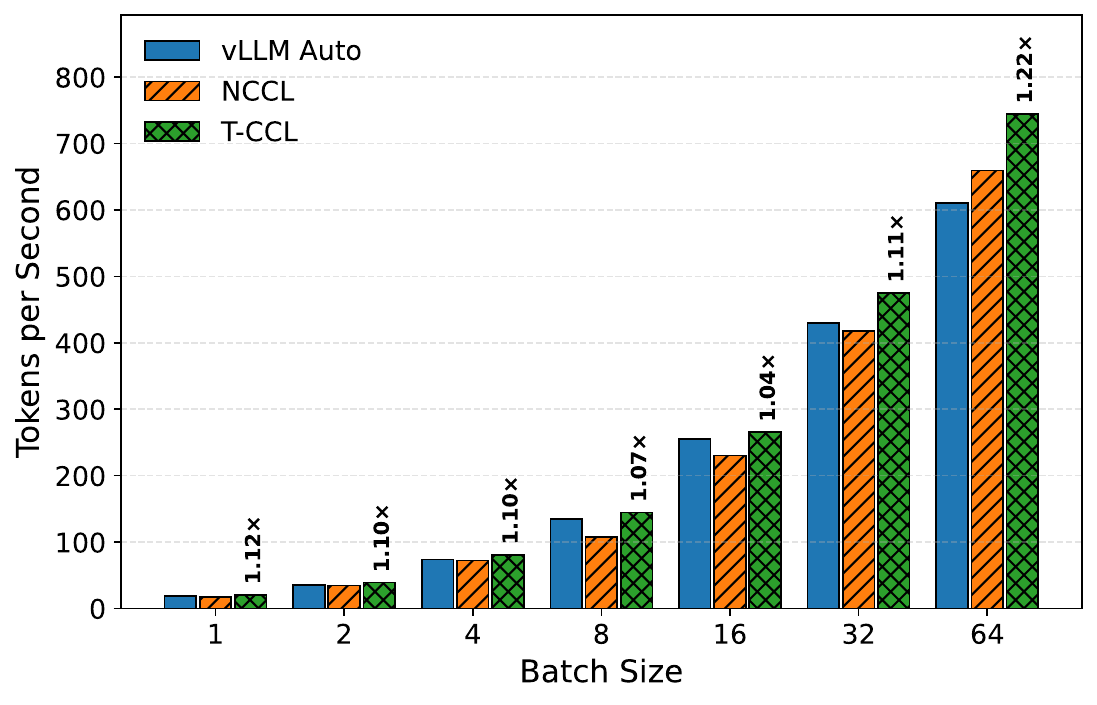}
        \label{fig:vllm-conversation}
    }%
    \hfill
    \subfloat[Decode-heavy workload: 512 input tokens and 1024 output
    tokens.]{  \includegraphics[width=0.48\textwidth]
        {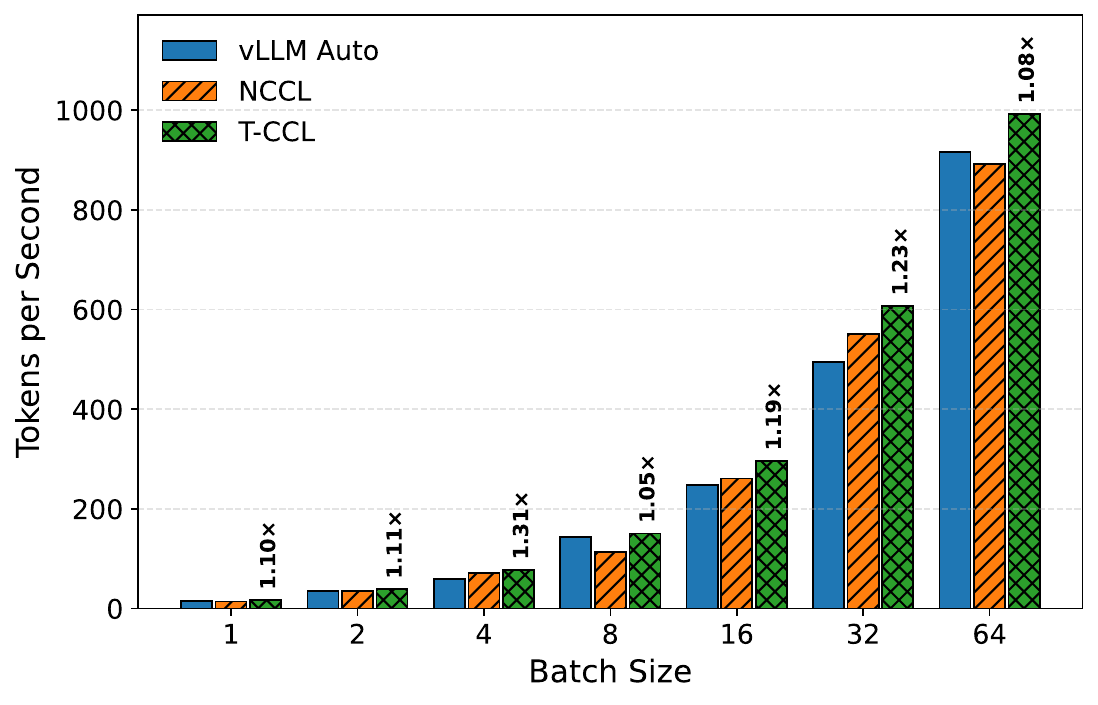}
        \label{fig:vllm-long-decode}
    }
    \caption{Throughput measurements for the conversation and decode-heavy vLLM
    workloads using the \textit{Qwen2.5-72B-Instruct} model with TP=4 on the
    GH200 platform. Results are shown for vLLM Auto, NCCL, and T-CCL. Labels
    above the T-CCL bars indicate speedup relative to vLLM Auto.}
    \label{fig:vllm-throughput}
\end{figure*}

\subsection{Case Study: Improving Compute--Communication Overlap}

We conduct a case study to evaluate T-CCL when GEMM and collective communication execute concurrently. 
We consider this setting because GEMM and collective communication frequently occur together in tensor-parallel transformer layers, and it allows us to assess whether T-CCL can further hide communication latency while limiting interference with computation.
Rather than waiting for the entire GEMM operation to finish, partial GEMM results are communicated as soon as they become ready while the remaining portions are still being computed. 
The size of each communicated segment, and therefore the granularity of overlap, is determined by the overlap algorithm and its configuration.
%
We use FlashOverlap~\cite{flashoverlap} as the overlap framework. 
FlashOverlap tracks GEMM output completion at tile granularity and groups
completed tiles into contiguous regions that can be communicated before the
GEMM finishes. Its signaling and reordering mechanisms keep the GEMM and
collective communication kernels separate. This separation allows the same GEMM
implementation and overlap schedule to be paired with different communication
backends, enabling a controlled comparison between NCCL and T-CCL.
The original FlashOverlap implementation targeted pre-Hopper GPUs and used
CUTLASS~2.x-style APIs~\cite{flashoverlap,cutlass_3x_gemm_api}. We ported FlashOverlap
to the Hopper architecture and CUTLASS~3.x APIs.

\paragraph{Configuration settings}
We compare three configurations:  cuBLAS~\cite{nvidia_cublas} followed by NCCL, representing sequential GEMM and communication (denoted as the cuBLAS--NCCL baseline),  and FlashOverlap using either NCCL or T-CCL as its communication backend (the NCCL variant and the T-CCL variant, respectively). We report the average speedup of the two FlashOverlap configurations over the sequential baseline across 24 GEMM shapes, with $M \times N$ ranging from $64 \times 1024^2$ to $384 \times 1024^2$ and $K \in \{2048, 4096, 8192\}$. 
We evaluate FlashOverlap configurations with both TP=2 and
TP=4 on the GH200 platform to maintain a consistent hardware environment, as the H100 and GH200 platforms have different numbers of SMs.

FlashOverlap communicates partial GEMM results in groups of output tiles.
To maximize overlap, it uses a search algorithm to select the best group size
based on the available communication resources (CTAs) and the bandwidth of
the communication backend~\cite{flashoverlap}. Therefore, for TP=2, we
provide CTA budgets of 2, 4, 6, 8, 10, 12, and 16 as the available
communication resources for both T-CCL and NCCL. For TP=4, we provide CTA
budgets of 3, 6, 9, 12, 15, 18, and 24. We evaluate all configurations
considered by the search algorithm and select the one with the lowest execution
time.

\paragraph{Overlap speedup}
Fig.~\ref{fig:operator-overlap-speedup} shows the average speedup of
FlashOverlap with NCCL and T-CCL relative to the sequential baseline (cuBLAS--NCCL), together with
the minimum and maximum speedups observed across the 24 GEMM shapes.
Overlap improves performance in both settings. For TP=2, the NCCL and T-CCL variants achieve average speedups of 1.12$\times$ and
1.25$\times$, respectively, with the T-CCL variant reaching up to 1.51$\times$ for the shapes that offer the most computation to overlap. 
For TP=4, they achieve 1.04$\times$ and 1.14$\times$, with a maximum of 1.34$\times$ for the T-CCL variant. 
In both cases, T-CCL provides a higher average speedup
because its smaller SM-side resource footprint leaves more resources
available for GEMM computation.

For TP=2, our T-CCL variant maintains a speedup above the sequential baseline across all 24 GEMM shapes, whereas the NCCL variant falls below the baseline in its worst case.
For TP=4, where each GPU holds a smaller share of computation, both variants exhibit minimum speedups below the sequential baseline. These worst cases generally occur for smaller values of $M$, $N$, and $K$: small GEMMs provide little computation with which to overlap communication, so the configuration selected by FlashOverlap's predictive search may perform worse than simply running GEMM and communication sequentially. 
The remaining shapes retain a speedup above the sequential baseline, with the largest gains on compute-rich shapes. Overall, these results indicate that in overlap scenarios, the small resource footprint of T-CCL makes it a promising communication backend, delivering higher average speedup than NCCL at both TP degrees.

\subsection{Case Study: Improving LLM Inference Performance}

We next evaluate T-CCL in an LLM inference setting to determine whether its communication-level improvements also translate into higher end-to-end inference performance. 
We use vLLM~\cite{vllm} and integrate T-CCL as a communication backend supporting AllReduce, ReduceScatter, and AllGather. We compare three configurations: \textit{vLLM Auto}, which retains vLLM's
built-in communication dispatch among its custom AllReduce, NCCL with
symmetric memory, and standard NCCL; \textit{NCCL}, in which the evaluated
collectives are forced to use NCCL; and \textit{T-CCL}, in which they use
T-CCL. All experiments use the \textit{Qwen2.5-72B-Instruct} model~\cite{qwen} with
TP=4 on the GH200 platform. 

We run all experiments using vLLM's internal benchmarking utility~\cite{vllm}
with CUDA Graph execution disabled (i.e., in eager mode), since
T-CCL does not currently support CUDA Graph execution. 
CUDA Graphs reduce repeated host-side launch overhead by replaying captured GPU operations~\cite{cuda_guide}.
We therefore expect T-CCL to benefit from CUDA Graph support, as part of its collective
planning and launch orchestration is performed on the host. NCCL similarly benefits from
graph execution because it also incurs host-side launch overhead, while the GPU-side
communication differences between T-CCL and NCCL remain. We leave the development of
CUDA Graph support for T-CCL and its end-to-end evaluation in vLLM to future work.

%
We evaluate two workloads. The first uses an input length of 1024 tokens
and an output length of 128 tokens, closely matching the median input and
output lengths observed in a production LLM conversation trace~\cite{splitwise}.
The second uses input and output lengths of 512 and 1024 tokens,
respectively, representing a decode-heavy workload~\cite{nanoflow}. 
For both workloads, we vary the maximum number of active sequences (batch size) from
1 to 64 and report end-to-end output throughput.

Fig.~\ref{fig:vllm-throughput}\subref{fig:vllm-conversation} and~\ref{fig:vllm-throughput}\subref{fig:vllm-long-decode}
present the results. For the conversation workload, T-CCL achieves
geometric-mean speedups of 1.10$\times$ over vLLM Auto and 1.17$\times$
over NCCL across the evaluated settings, while outperforming both
configurations in every case and reaching up to 1.22$\times$ over vLLM Auto.
For the decode-heavy workload, T-CCL achieves geometric-mean speedups of
1.15$\times$ over both vLLM Auto and NCCL, again outperforming both
configurations throughout the evaluation, with a maximum speedup of
1.31$\times$ over vLLM Auto. The larger speedup over vLLM Auto on the
decode-heavy workload comes from the longer output sequence, which requires
more decode iterations and repeated collective invocations.
These results show that the communication-level advantages of T-CCL can
translate into higher end-to-end inference throughput for both representative
conversation and decode-heavy inference workloads.


\section{Conclusion}
This paper presented T-CCL, a TMA-based collective communication library
for intra-node GPU systems. T-CCL offloads the data movement and reduction
operations required by collectives to TMA. It translates AllReduce,
ReduceScatter, and AllGather into a series of logical tasks, while each
rank lowers these tasks into physical transfer operations that are
efficiently executed by a pipelined TMA engine.
Our evaluation showed that T-CCL generally achieves higher performance than NCCL with lower resource usage. 
Two case studies demonstrated the practical benefits of T-CCL: in compute--communication overlap, the SM resources freed by T-CCL yield greater overlap speedup, and integrating T-CCL into vLLM translates its gains on individual collectives into higher end-to-end inference throughput.

For future work, we plan to support window-level completion signaling to
more closely pipeline dependent copy and reduction operations. 
We will also extend T-CCL to support All-to-All for
mixture-of-experts workloads, where communication patterns and resource
demands can vary dynamically across GPUs. 
Finally, we plan to evaluate T-CCL on larger GPU topologies, such as
8- and 16-GPU systems, and investigate topology-aware planning for
asymmetric systems, where GPUs may be connected through interconnects
with heterogeneous bandwidth and latency characteristics.



\section*{Acknowledgment}
This work was supported by the Swedish Foundation for Strategic Research (SSF) under the PRIDE project (ref. no. CHI19-0048) and the Swedish Research Council (VR) under the P4PIM project (grant ID 2020-04892).
Computational resources were provided by the National Academic
Infrastructure for Supercomputing in Sweden (NAISS), under project
No.~2026/4-335, funded by the Swedish Research Council.
The computations were also enabled by resources provided by
Chalmers e-Commons at Chalmers.

Throughout the writing of this paper, across all sections, the authors used Grammarly~\cite{grammarly} and
ChatGPT~\cite{chatgpt} to improve the clarity of the paper. This included grammar
and vocabulary checking, as well as paraphrasing sentences originally
written by the authors. The authors then reviewed, edited, and rephrased
the suggested sentences in their own words. Therefore, the authors take
full responsibility for the content of the paper.







\IEEEtriggeratref{22}
\bibliographystyle{IEEEtran}
\bibliography{references}

\clearpage
\appendices
\appendixAD
\label{app:artifact-description}

\section{Overview of Contributions and Artifacts}

\subsection{Paper's Main Contributions}

\begin{description}

\item[\(C_1\)]
We design and implement T-CCL, an intra-node collective
communication library that uses the Tensor Memory Accelerator
for communication and reduction between GPUs. T-CCL supports
AllReduce, ReduceScatter, and AllGather.

\item[\(C_2\)]
We evaluate T-CCL against NCCL and symmetric NCCL for
intra-node collective communication. We characterize its
performance under unrestricted and restricted CTA budgets.

\item[\(C_3\)]
We evaluate T-CCL's performance benefits in a fine-grained
compute--communication overlap setting using FlashOverlap.

\item[\(C_4\)]
We integrate T-CCL into vLLM as a tensor-parallel communication
backend and evaluate its effect on end-to-end LLM inference
throughput.

\end{description}

\subsection{Computational Artifacts}

\begin{description}

\item[\(A_1\)]
The T-CCL artifact package contains the source code,
build system, environment setup scripts, benchmark drivers,
experiment configurations, raw-result processing tools, and
plotting scripts.

The artifact is available at:
\begin{center}
\url{https://github.com/ooverlap/ooverlap}
\end{center}
Artifact branch:
\texttt{anonymouse}.\\
Artifact version:
\texttt{9abd9cf}.

\end{description}

\begin{table*}[t]
\centering
\caption{Relationship between the paper's contributions,
computational artifacts, and paper elements supported by the artifact.}
\label{tab:artifact-mapping}
\small
\begin{tabularx}{\textwidth}{
    >{\centering\arraybackslash}p{0.09\textwidth}
    >{\centering\arraybackslash}p{0.15\textwidth}
    X}
\toprule
Artifact &
Contributions &
Paper elements supported by the artifact \\
\midrule

\(A_1\) &
\(C_1\) &
T-CCL implementation and system design described in
Section~III and illustrated in Figs.~1 and~2. \\

\(A_1\) &
\(C_2\) &
Standalone collective evaluation in Section~IV-B, including
Figs.~3--4 and the corresponding performance and SM-usage tables. \\

\(A_1\) &
\(C_3\) &
Compute--communication overlap evaluation using FlashOverlap in
Section~IV-C, including Fig.~5. \\

\(A_1\) &
\(C_4\) &
vLLM integration and end-to-end inference evaluation in
Section~IV-D, including Fig.~6. \\

\bottomrule
\end{tabularx}
\end{table*}

\section{Artifact Identification}

\newartifact

\artrel

Artifact \(A_1\) supports contributions \(C_1\)--\(C_4\).

For \(C_1\), the artifact contains the complete T-CCL
implementation and the scripts required to configure, build, and
test its AllReduce, ReduceScatter, and AllGather operations.

For \(C_2\), the artifact contains the benchmark drivers,
experiment configurations, result-processing tools, and plotting
scripts used for the standalone collective evaluation.

For \(C_3\), the artifact contains the ported FlashOverlap source code,
together with the scripts for running and plotting the
compute--communication overlap experiments.

For \(C_4\), the artifact contains the T-CCL integration for
vLLM and the scripts used to execute and analyze the
tensor-parallel inference experiments.

\artexp

Successful reproduction of the standalone collective experiments
should generate latency and algorithmic bandwidth measurements
for T-CCL, NCCL, and NCCL's symmetric-memory kernels under unrestricted
and restricted CTA budgets. These results support \(C_2\) by reproducing
the collective-performance trends reported in the paper.

Successful reproduction of the overlap experiments should generate
execution-time and speedup results for the sequential cuBLAS--NCCL
baseline and FlashOverlap using NCCL or T-CCL. These results support
\(C_3\) by reproducing the performance benefit of T-CCL in
compute--communication overlap.

Successful reproduction of the vLLM experiments should generate
output-throughput measurements for vLLM Auto, NCCL, and T-CCL across the
evaluated maximum numbers of active sequences. These results support
\(C_4\) by reproducing the end-to-end inference throughput improvements
reported in the paper.

The reproduced results should exhibit the same general performance trends
as Figs.~3--6 and Tables~I, III, and IV. Exact execution times may
differ because of GPU clocks, machine topology, thermal conditions, and system load.

\arttime

The approximate reproduction times are:

\begin{itemize}
    \item Artifact setup and compilation: approximately
          \textbf{10 minutes}.
    \item Artifact execution:
          approximately \textbf{20 minutes} for the complete standalone
          collective evaluation, approximately \textbf{300 minutes per TP
          configuration} for the compute--communication overlap evaluation,
          and approximately \textbf{360 minutes} for the vLLM evaluation.
    \item Artifact analysis and plot generation are performed automatically
            by the evaluation scripts and are included in the execution times above.
\end{itemize}

Execution times may vary depending on the hardware, system load, and software environment.

\artin

\artinpart{Hardware}

The paper evaluates T-CCL on the following systems:

\begin{itemize}
    \item A two-GPU system containing two NVIDIA H100 NVL GPUs,
          connected through NV12.
    \item A four-GPU system containing four NVIDIA GH200 120-GB
          GPUs, connected through NV6.
\end{itemize}

The implementation targets NVIDIA Hopper GPUs with SM90 support.
It should require only minor build changes for newer NVIDIA GPUs
with TMA support. Reproducing the reported performance
requires hardware and interconnects similar to the evaluated systems.

\artinpart{Software}

The principal software dependencies and versions are:

\begin{itemize}
    \item Linux kernel
          \texttt{5.14.0-687.26.1.el9\_8.aarch64+64k}
          on the GH200 system and
          \texttt{5.14.0-687.30.1.el9\_8.0.1.x86\_64}
          on the H100 system.

    \item NVIDIA driver 580.159.04 on the GH200 system and
          595.71.05 on the H100 system
          {\footnotesize\url{https://www.nvidia.com/Download/index.aspx}}.

    \item CUDA Toolkit 13.0 on the GH200 system and
          CUDA Toolkit 13.3 on the H100 system
          {\footnotesize\url{https://developer.nvidia.com/cuda-toolkit-archive}}.

    \item NVIDIA Nsight Systems 2025.3.2.

    \item NCCL 2.28
          {\footnotesize\url{https://github.com/NVIDIA/nccl/releases/tag/v2.28.9-1}}.

    \item NCCL 2.30.7
          {\footnotesize\url{https://github.com/NVIDIA/nccl/releases/tag/v2.30.7-1}}
          with an artifact-provided patch enabling the TMA path on Hopper
          for the SM-usage experiments.

    \item CMake 4.4.0
          {\footnotesize\url{https://cmake.org/cmake/help/v4.4/release/4.4.html}}.

    \item GCC 14.3.0
          {\footnotesize\url{https://gcc.gnu.org/gcc-14/}}.

    \item Python 3.12 or newer
          (Python 3.13.5 was used on both systems;
          {\footnotesize\url{https://www.python.org/}}).

    \item PyTorch \texttt{2.11.0+cu130}
          {\footnotesize\url{https://pytorch.org/get-started/previous-versions/}}.

    \item vLLM 0.25.0
          {\footnotesize\url{https://github.com/vllm-project/vllm/releases/tag/v0.25.0}}
          with an artifact-provided patch for T-CCL integration,
          applied automatically during setup.

    \item CUTLASS 3.9.0
          {\footnotesize\url{https://github.com/NVIDIA/cutlass/releases/tag/v3.9.0}}
          with an artifact-provided patch for the FlashOverlap evaluation,
          applied automatically during setup.
\end{itemize}

\artinpart{Datasets / Inputs}

The standalone collective experiments use synthetically generated
GPU buffers and do not require an external dataset.

The compute--communication overlap experiments require no external
dataset. They use synthetically generated FP16 input matrices,
while the matrix dimensions, communication configurations, and
evaluation parameters are defined by the artifact.

The vLLM experiments use the built-in random-data benchmark to
generate synthetic prompts. All vLLM experiments use the
\texttt{Qwen/Qwen2.5-72B-Instruct} model with tensor-parallel
size four. The exact model revision is
{\footnotesize\texttt{495f39366efef23836d0cfae4fbe635880d2be31}}.

The evaluated workloads are:

\begin{itemize}
    \item Conversation workload: 1024 input tokens and
          128 output tokens.
    \item Decode-heavy workload: 512 input tokens and
          1024 output tokens.
    \item Maximum active sequences: 1, 2, 4, 8, 16, 32, and 64.
\end{itemize}

\artinpart{Installation and Deployment}

The artifact must be cloned with all Git submodules:

\begin{artifactcommand}
git clone --recursive https://github.com/ooverlap/ooverlap
cd ooverlap/
\end{artifactcommand}

The artifact provides separate setup scripts for the two evaluated
platforms:

\begin{artifactcommand}
bash setup_env.sh [venv-path]       # H100/x86
bash setup_arm_env.sh [venv-path]   # GH200/ARM
\end{artifactcommand}

By default, the scripts load the site-specific environment modules
used on the evaluated systems. On machines with an equivalent
compiler and CUDA toolchain, module loading can be disabled and a
custom virtual-environment location can be specified:

\begin{artifactcommand}
SKIP_MODULES=1 bash setup_env.sh /path/to/venv
\end{artifactcommand}

This module-independent setup has been validated on x86 systems but
has not yet been validated on ARM systems. The scripts install the
required Python packages, configure the patched CUTLASS dependency,
apply the vLLM integration patch, and build T-CCL and its PyTorch
extension.

After setup, the script prints the path of a generated environment
file. This file must be sourced before running the evaluation:

\begin{artifactcommand}
source /path/to/ooverlap_vllm_env.sh
\end{artifactcommand}

\artcomp

All workflows require T-CCL and its Python extensions to be built
beforehand, and the generated runtime-environment file to be sourced.
The three workflows can then be executed independently. Optional tuning \(T_0\) produces the launch-configuration data used
by T-CCL. The standalone collective workflow consists of \(T_1\) and
\(T_{\mathrm{SM}}\), the overlap workflow consists of \(T_2 \rightarrow T_3\),
and the vLLM workflow consists of \(T_4 \rightarrow T_5\).

\paragraph{Optional T-CCL Tuning}

\begin{description}

\item[\(T_0\)]
From the T-CCL repository root, run:

\begin{artifactcommand}
python3 test/sweep_tma_collectives.py \
    --devices <device-list>
\end{artifactcommand}

The device list must contain either two or four GPU device identifiers,
corresponding to TP=2 or TP=4, respectively. This script evaluates
different T-CCL launch configurations and generates the tuning results.
This step is not required to run the evaluation workflows below.

\end{description}

\paragraph{Standalone Collective Evaluation}

\begin{description}

\item[\(T_1\)]
Run:

\begin{artifactcommand}
cd evalution
./run_external_p2p_bandwidth_latency.sh <tp> <mode>
\end{artifactcommand}

The first argument is either \texttt{2} or \texttt{4}. The second
argument is either \texttt{unres} for the unrestricted CTA
configuration or \texttt{res} for the restricted CTA configuration.
Run all four combinations to reproduce the complete standalone
evaluation. Each invocation automatically generates the corresponding
plots and reports the minimum and maximum T-CCL speedups over NCCL and
symmetric NCCL for both the small- and large-message ranges across all
collectives.

\item[\(T_{\mathrm{SM}}\)]
To measure the average active SM usage, run:

\begin{artifactcommand}
cd evalution
./run_external_p2p_sm_usage.sh <tp> <op> <numel> [<numel> ...]
\end{artifactcommand}

The first argument is either \texttt{2} or \texttt{4}, corresponding
to TP=2 or TP=4. The second argument specifies the collective operation:
\texttt{AG} for AllGather, \texttt{RS} for ReduceScatter, or
\texttt{AR} for AllReduce. The remaining arguments specify one or more
tensor sizes as numbers of elements. The script measures and reports the
average active SM usage of T-CCL, NCCL, symmetric NCCL, and NCCL with
TMA enabled for each specified tensor size.

\end{description}

\paragraph{Compute--Communication Overlap Evaluation}

\begin{description}

\item[\(T_2\)]
Build the patched CUTLASS profiler, copy
\path{scripts/sweep_h100_dense_f16_gemm_same_grid.sh}
into the CUTLASS profiler directory, and run it to generate the
required GEMM profile CSV files.

\item[\(T_3\)]
From the T-CCL repository root, run:

\begin{artifactcommand}
cd evalution
./run_flashoverlap_sm90.sh <tp>
\end{artifactcommand}

The argument is either \texttt{2} or \texttt{4}. The script runs
the overlap evaluation and automatically generates the plots.

\end{description}

\paragraph{vLLM Evaluation}

\begin{description}

\item[\(T_4\)]
Download the model revision specified above and record its local
path.

\item[\(T_5\)]
Run the vLLM evaluation with tensor-parallel size four:

\begin{artifactcommand}
cd evalution
VLLM_MODEL_DIR=/path/to/model \
    ./run_vllm_paperlike.sh 4 run
\end{artifactcommand}

The script runs both the conversation and decode-heavy workloads
and automatically generates their corresponding plots.

\end{description}

\newpage
\artout

The standalone collective analysis calculates latency for small
messages and algorithmic bandwidth for large messages. Results
are grouped by collective operation, system, message size,
communication backend, and CTA budget.

The overlap analysis compares the execution time of
FlashOverlap with NCCL or T-CCL against the sequential
cuBLAS--NCCL baseline and calculates the resulting speedup.

The vLLM analysis groups successful runs by workload,
communication backend, and maximum number of active sequences
(batch size). It then computes output-token throughput and the
relative speedup of T-CCL over the evaluated baselines.

All reproduced plots are generated in PNG or PDF format.


\end{document}